\documentclass[fleqn,usenatbib]{mnras}

\usepackage{newtxtext,newtxmath}

\usepackage[T1]{fontenc}

\DeclareRobustCommand{\VAN}[3]{#2}
\let\VANthebibliography\thebibliography
\def\thebibliography{\DeclareRobustCommand{\VAN}[3]{##3}\VANthebibliography}

\usepackage{graphicx}	
\usepackage{amsmath}	

\newcommand{\Msolar}{\mbox{\,$\rm M_{\odot}$}}        

  \newcommand{\Teff}{\mbox{\,\em T$_{\rm eff}$}}         

  \newcommand{\degrees}{\mbox{$^\circ$}}                 
  \def\simge{\mathrel{\raise1.16pt\hbox{$>$}\kern-7.0pt
    \lower3.06pt\hbox{{$\scriptstyle \sim$}}}}           
  \def\simle{\mathrel{\raise1.16pt\hbox{$<$}\kern-7.0pt
    \lower3.06pt\hbox{{$\scriptstyle \sim$}}}}           
\title[Surface features on HD\,49798]{Surface brightness inhomogeneity on the helium-rich hot subdwarf in the X-ray binary HD\,49798}

\author[C. S. Jeffery \& G. Ramsay]{
C. Simon Jeffery\thanks{E-mail: simon.jeffery@armagh.ac.uk} and
Gavin Ramsay
\\
The Armagh Observatory and Planetarium, College Hill, Armagh BT61 9DG, United Kingdom
}

\date{Accepted XXX. Received YYY; in original form ZZZ}

\pubyear{\the\year{}}

\begin{document}
\label{firstpage}
\pagerange{\pageref{firstpage}--\pageref{lastpage}}
\maketitle

\begin{abstract}
HD\,49798 is an X-ray binary consisting of a hot helium-rich subdwarf primary and an  accreting compact companion with an orbital period of 1.55\,d. 
TESS observations of its optical light curve suggest short-period brightenings superimposed on a sinusoidal variation.
The latter can be attributed to the tidal deformation of the primary, while frequency analysis of the additional variations shows up to eleven frequencies which form an harmonic series based on a period of 1.4\,d, in excellent agreement with the stellar rotation period inferred independently from spectroscopy. 
These brightenings may be explained by a slowly changing and inhomogeneous temperature distribution across the surface of the hot subdwarf, such as might be associated with a magnetic field or with the presence of Rossby waves in the stellar envelope.  
\end{abstract}




\section{Introduction}

HD\,49798 is a well-studied bright low-mass X-ray binary  consisting of a hot subdwarf with a compact companion in a 1.55 day orbit.
The hot subdwarf is a high-mass and helium-enriched sdO star \citep{rauch25}. 
Whether the companion is a white dwarf or neutron-star is not yet certain \citep{Brooks2017}.
X-rays originate from a magnetized hot spot on the compact star and indicate a spin-period of 13.2s \citep{Israel1996}.
Since the X-rays are eclipsed by the sdO star, the orbital inclination must be high \citep{mereghetti16b}. 
There has been no report of an eclipse at visible wavelengths. 
Light curves of hydrogen-rich hot subdwarfs in binary systems observed with the Terrestrial Exoplanet Sky Survey (TESS) show evidence of ellipsoidal deformation and, where present, non-radial oscillations; some show eclipses \citep{barlow22}. 
In contrast, detections of variability in helium-rich hot subdwarfs are rare \citep{snowdon25}.
Where present, variability may point to evidence of non-radial pulsations \citep[{\it e.g.} LS\,IV$-14^{\circ}116$:][]{ahmad05b}, a close companion \citep[{\it e.g.} Ton\,S\,415:][]{snowdon23b},  or be of unknown origin \citep[{\it e.g.} BPS CS 22956-0094, EC\,21077-4815:][]{snowdon25}. We have therefore, from time to time, reviewed TESS light curves for helium-rich hot subdwarfs, including HD\,49798. 

\begin{figure}
	\includegraphics[width=\columnwidth,clip, trim=0 0 0 0]{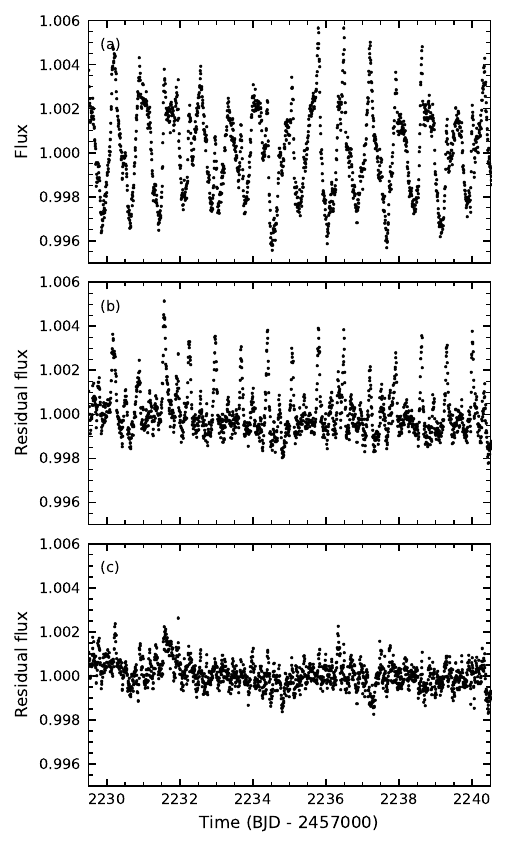}
    \caption{Panel (a): part of the TESS sector 34 light curve for HD\,49798 normalized to the mean flux. Panel (b) shows the same data with the ellipsoidal component removed.  Panel (c) shows the same data with both the ellipsoidal and additional components removed. }
    \label{fig:lightcurve}
\end{figure}

\section{Observations}
HD\,49798 (TIC 170203297) was observed with TESS in sectors 6 and 7 (2018-12-12 to 2019-02-01) at a cadence of 120 s and  sectors 33 and 34 (2020-12-18 to 2021-02-08) at a cadence of 20 s.
Reduced PDC (Pre-search Data Conditioning) light curves were downloaded from the Mikulski Archive of Space Telescope (MAST) data with the light curve analysis package {\sc lightkurve}  \citep{lightkurve18}. 
Part of the  light curve for sector 33 is shown in Fig.\,\ref{fig:lightcurve}a.  
It suggests a sinusoidal varation on roughly the orbital period as well as higher frequency variations. 
Since the system is a known X-ray binary, the question arises whether the light curve contains additional information about the companion. 

\begin{figure}
	\includegraphics[width=\columnwidth,clip, trim=0 0 0 0]{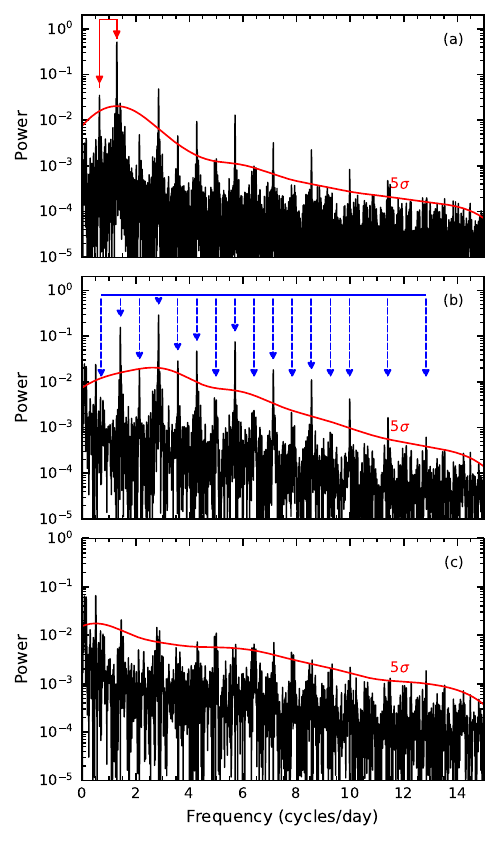}
    \caption{Panel (a): The Lomb-Scargle power spectrum for the entire TESS light curve of HD\,49798. The position of the orbital frequency and its first harmonic are marked by tied red arrows.  The 5$\sigma$ confidence level is shown by a smooth red line. Panel (b): As (a) for sectors 33 and 34 pre-whitened by the ellipsoidal variation. The positions of the additional frequency and  harmonics are marked by tied blue arrows. Panel (c): As (b) pre-whitened by both the ellipsoidal and additional modulation.  }
        \label{fig:power}
\end{figure}

\begin{table}
	\centering
	\caption{Frequencies $f$ and amplitudes $a$ present in the TESS light curve for HD\,49798. The orbit-related frequencies are obtained from all four sectors. The  additional frequencies are obtained from sectors 33 and 34 (left), with $f_1 \equiv f_4 / 4$  and from the noisier sectors 6 and 7 (right).  $\delta$ represents the difference between the measured frequency $f_n$ and the harmonic frequency $nf_1$. Periods $P$ are shown for the fundamental frequency $f_1$ in each series.  }
	\label{tab:frequencies}
	\begin{tabular}{rrrrrrr} 
		\hline
		$n$ & $f_n$ (${\rm d}^{-1}$) & $\delta$ & $a$ (\%) & $P$ (d) & $f_n$ (${\rm d}^{-1}$) & $a$ (\%) \\
		\hline
\multicolumn{5}{l}{Orbit (all sectors)} & \\
1 & 0.64603 &  0.0000 & 0.0563 & 1.5511 \\
2 & 1.29229 & -0.0003 & 0.3661 &        \\[1mm]
\multicolumn{5}{l}{Additional (sectors 33+34)}& \multicolumn{2}{l}{(sectors 6+7)}\\
1 & (0.7123) &      &        & 1.4038 & 0.72705 & 0.0497 \\
2 & 1.42225 & 0.0024 & 0.0601 &  & 1.44310 &0.0441\\
3 & 2.13399 & 0.0030 & 0.0197 &  & \\
4 & 2.84934 & 0.0000 & 0.0821 &  & 2.85619 & 0.0590 \\
5 & 3.57037 & -0.0087 & 0.0268 & & 3.48923 & 0.0308 \\
6 & 4.27474 & -0.0007 & 0.0324 & & 4.29028 & 0.0325 \\
7 & 5.00415 & -0.0178 & 0.0106 & & \\
8 & 5.70429 & -0.0056 & 0.0417 & & 5.73638 & 0.0251\\
10 & 7.12874 & -0.0054 & 0.0207 & & \\ 
12 & 8.55674 & -0.0087 & 0.0162 & & \\
14 & 9.98068 & -0.0080 & 0.0099 & & \\
16 & 11.40970 & -0.0123 & 0.0058 & & \\
		\hline
	\end{tabular}
\end{table}

\section{Analysis}

The TESS data were analysed for sectors 6 and 7 and sectors 33 and 34 separately, and for all four sectors together. 
Lomb-Scargle power spectra show multiple peaks (Fig.\,\ref{fig:power}a).  
The strongest and third strongest signals lie at frequencies of 1.29229 d$^{-1}$ and  0.64603 d$^{-1}$, respectively (Table 1).
 Both exceed the 5$\sigma$ confidence level by a substantial margin. 
The latter corresponds exactly with the orbital period $P_{\rm orb}\approx1.55$ d \citep{Thackeray1970}, whilst the former is its $n=2$ harmonic. Their amplitudes are in the ratio $0.366\%/0.056\% \sim 6.5$ and indicate an ellipsoidal variation \citep[cf.][]{green23}. An additional comb of significant frequencies is clearly present in the power spectrum, but not at harmonics of the orbital frequency.

\begin{figure}
	\includegraphics[width=\columnwidth,clip, trim=0 0 0 0]{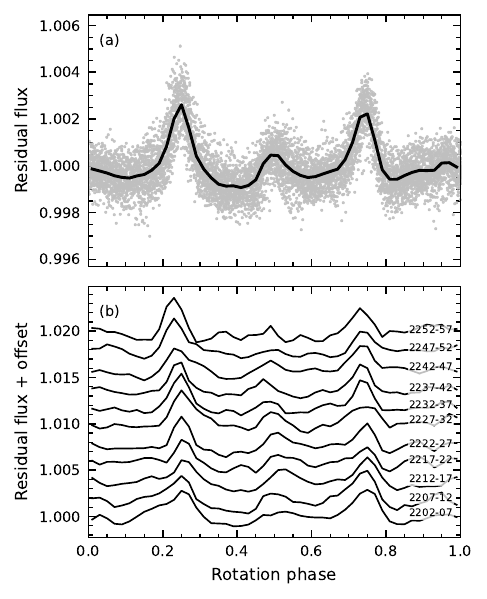}
    \caption{Panel (a): The residual light curve for HD\,49798 for sectors 33 and 34 folded on the  additional period of 1.4038 d. The bold line indicates the phase-binned average. Panel (b): The phase-binned average light curve for sectors 33 and 34 broken down into 5-day segments as labelled by BJD-2457000.}
    \label{fig:folded}
\end{figure}

\begin{figure}
	\includegraphics[width=\columnwidth,clip, trim=0 0 0 0]{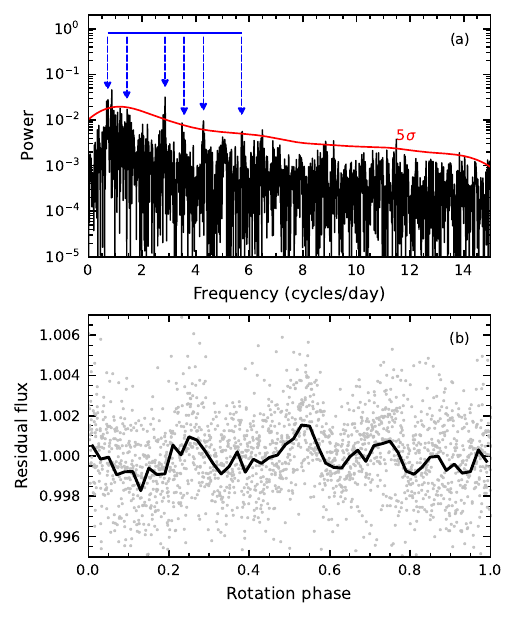}
    \caption{Panel (a): The power spectrum of the pre-whitened sector 6 and 7 light curve of HD 49798 (cf. Fig. 2(b)). The blue arrows mark the rotation harmonics identified in Table 1.  Panel (b): the same lightcurve folded on the additional period of 1.4038\,d with the phase-binned average in bold (cf. Fig. 3a).}
    \label{fig:sec67}
\end{figure}

The light curves were pre-whitened by the ellipsoidal variation using the period analysis package {\sc pywhiten} \citep{Pywhiten2023}. 
Part of the residual light curve is shown in Fig.\,\ref{fig:lightcurve}b. 
The Lomb-Scargle power spectra of the residual light curve continues to show regularly-spaced peaks with the strongest at a frequency of 2.84934 d$^{-1}$, with successive peaks showing alternating amplitudes (Fig.\,\ref{fig:power}b).   
The simplest interpretation  is that the stronger sequence corresponds to even-numbered harmonics and the weaker sequence to odd-numbered harmonics,  with the fundamental  ($f_1 = 0.7123$ d$^{-1}$) being unobserved. In other words, the waveform of the $f_1$ signal is highly non-sinusoidal with most power in higher harmonics.
Table 1 summarises the frequencies and amplitudes measured from sectors 33 and 34.
The residual light curve folded on a period of $P=f_1^{-1}=1.4038$ d is shown in Fig.\,\ref{fig:folded}a. 
 Similar behaviour is observed in the noisier data of sectors 6 and 7, despite the four year separation, providing evidence against instrumental origin. 

An additional stage of pre-whitening to remove both ellipsoidal and additional variations was conducted to establish whether the eclipse of the subdwarf by the white dwarf could be detected. 
With radii and orbital elements from \citet{rigoselli23}, the eclipse depth and duration would be around 20 parts per million (ppm) and 0.077 orbital cycles respectively. 
Phasing the residual light curve from Sectors 33 and 34 over the orbital period and binning to 0.005 cycles per orbit, the residual light curve has an rms and peak-to-peak variation of 94 ppm and 540 ppm respectively, yielding no evidence of an eclipse.

\section{Interpretation}

Estimates of the flux amplitude due to ellipsoidal deformation and Doppler beaming may be obtained from \citet[Eq. 15]{jackson12}.  
\citet{rigoselli23} report the orbital inclination $i\approx 84.5\degrees$, semi-major axis $a=0.0193$AU, eccentricity $\approx 0$, masses 
$M_{\rm sdO}\approx1.41\Msolar$ and $M_{\rm X}\approx1.22\Msolar$.  
\citet{rauch25} give the subdwarf radius $R_{\rm sdO}\sim1.1\,R_{\odot}$. 
Letting constants of order unity $= 1$, and supposing 
the mass ratio  $q=M_{\rm X}/M_{\rm sdO}\approx 0.8 - 1.0$, 
then the amplitude of the ellipsoidal variation 
$A_{\rm ellip}\sim 0.014 - 0.018$.
With radial-velocity semi-amplitude $K=118\,{\rm km\,s^{-1}}$, the Doppler beaming amplitude 
$A_{\rm beam}\simeq1.57\times10^{-3}$ or $\sim A_{\rm ellip}/10$.
These simple estimates are within an order of magnitude of amplitudes 0.00366 and 0.00056 measured for the first harmonic and fundamental terms in the orbital light curve of HD\,49798 (Table\,\ref{tab:frequencies}). 

A significant observation is that the underlying $P = 1.4$\,d for the remaining photometric variation is almost exactly the same as the rotational period measured spectroscopically.  
\citet{rauch25} find the projected rotation velocity $v_{\rm rot}\sin i  \sim40\,{\rm km\,s^{-1}}$. 
Since the inclination $i\approx 84.5\degrees$, the rotation period, which is not derived from photometry, is simply $P_{\rm rot} = 2 \pi R \sin i/v_{\rm rot} \sin i \approx1.4$ d.
 The detection does not rely on a single Fourier peak but on eleven harmonically related frequencies whose deviations from exact integer multiples are below 0.01 d$^{-1}$. 
The close agreement between the inferred rotational period and the photometric modulation period strongly supports a rotational origin for the residual variability.
 The light curve shown in Fig.\,\ref{fig:folded}a could be thus be due to localized bright regions on the surface of the sdO star, possibly associated with a structured magnetic field or with Rossby waves in the stellar atmosphere. 

 Fig.\,\ref{fig:folded}b shows the residual light curve during sectors 33 and 34 resolved into 5\,d time bins.  
Some bright peaks appear to drift in phase relative to others, indicative of evolving bright structures and imply a dynamic  field topology or wave pattern.

Figs.\,\ref{fig:lightcurve}c and \ref{fig:power}c show a section of the light curve and power spectrum after subtraction of both the orbital modulation and the additional harmonic series identified in Table 1. The persistence of some harmonics after subtraction of both signals is consistent with slow evolution of the waveform.
The lower-sampling rate of the sector 6 and 7 data makes it difficult to compare the rotational light curve over the full four-year baseline, but similar structures and frequencies are evident (Fig.\,\ref{fig:sec67}, Table 1).
  
Given that HD\,49798 is an X-ray binary, we conjectured that the residual light curve arises in a precessing tilted disk around the compact object. An elementary analysis shows that a classical accretion disk with a superhump period of 1.4\,d would be difficult to accommodate within the Roche geometry of the compact object. 
 A hot spot on the face of the subdwarf heated by X-rays could a produce a shallower minimum in the orbital light curve at inferior conjunction of the white dwarf (Schaffenroth et al. in prep.); a structured light curve at the rotation period would be unlikely.   
X-rays from the hot spot imply that material is accreted onto the compact companion, presumably from a weak subdwarf wind \citep{mereghetti21}; if some of this wind were reaccreted onto the subdwarf along magnetic field lines, for example, the subdwarf rotation period would manifest. 
Low-amplitude pulsations are seen in a few subdwarf O stars \citep{koen97,kilkenny97,woudt06,randall11} with periods on the stellar dynamical timescale of seconds to minutes and amplitudes of a per cent. 
These are usually identified with low-order p and g modes; a mode having a period $\sim 1.4$ d would require the star to be a giant. None of these mechanisms seem plausible explanations for the signal observed at harmonics of the rotation frequency.

\section{Discussion}

Confirmation that the additional signal at harmonics of the rotation frequency is not an artefact of prewhitening is given by its significant presence in the original power spectrum (Fig.\ref{fig:lightcurve}a). 
Since the TESS light curve is  well sampled, non-linear consequences of pre-whitening are expected to be small. 
To test whether residual leakage could introduce an artificial secondary waveform we constructed  an artificial  light curve using the timestamps of the TESS data. 
We injected an orbital modulation using the 2 orbital components from Table 1 and added incoherent shuffled noise with root-mean-square amplitude 0.0002. 
We injected a third component at $11f_{\rm orb}$ and then pre-whitened the light curve by the two known frequencies. The peak at $11f_{\rm orb}$ was unaffected and no additional peaks were created.  

\subsection{Magnetic fields}

Evidence of surface magnetic fields in hot subdwarf stars is sparse. 
\citet{Landstreet2012} found scant evidence for magnetic fields stronger than a few hundred Gauss in a sample of 40 hot subdwarfs with effective temperatures  in the range $T_{\rm eff} = 20\,000 - 100\,000$ K. 
The only positive detections to date are in sdO stars with  $T_{\rm eff} =  44\,000 - 48\,000$ K 
and surface helium-to-hydrogen ratios in the range $n_{\rm He}/n_{\rm H} = 0.5 - 3.0$ \citep{pelisoli22,dorsch22,dorsch24} 
where field strengths measured from the Zeeman effect are 200 — 500 kiloGauss. 
An additional spectral property of these magnetic He-sdOs is the presence of a broad absorption feature at 4630\AA. 
TESS light curves are not available for any of the strongly magnetic He-sdOs. 
One He-sdO (J1346-4025$\equiv$UCAC4 248-062759) which shows the 4630\AA\ absorption also shows a sinusoidal  modulation in its TESS light curve with a period of 5.4 d. 
This is commensurate with a rotational modulation, but does not necessarily imply a magnetic field. 

HD\,49798 has $T_{\rm eff} =  45\,000 \pm 1\,000 $ K and $n_{\rm He}/n_{\rm H} = 3.9 \pm 0.3 $ \citep{rauch25}.
With a surface gravity $\log g/{\rm cm\,s^{-2}} = 4.46\pm0.10$ (ibid.), it is more luminous than the known magnetic He-sdOs ($\log g/{\rm cm\,s^{-2}} = 5.5 - 6.1$: \citet{dorsch24}). 
An archival ESO/UVES spectrum shows no evidence of broad 4630\AA\ absorption, and no Zeeman effect has been reported.

The evidence that HD\,49798 and confirmed magnetic He-sdOs appear to occupy a small range in $T_{\rm eff}$ begs the question of the magnetic field origin.
\citet{dorsch24} argue for a dynamo-driven field consequent on differential-rotation following the merger of two white dwarfs. 
Being still a binary, HD\,49798 has a different history involving two or more mass-transfer episodes \citep{Brooks2017}. 
Two mechanisms proposed to explain rotation-modulated light curves of early O stars involve a magnetic field. One requires a large-scale fossil field \citep[e.g. $\sigma$ Ori E:][]{landstreet78}; the other requires a subsurface convection zone driven by the iron-group opacity bump to produce a small-scale field that is buoyed to the surface \citep{Cantiello2011}. 
Good examples include $\xi$ Per and $\zeta$ Pup \citep{Ramiaramanantsoa14,Ramiaramanantsoa2018}. 
Further modelling would be required to develop either hypothesis for HD\,49798. 

We note evidence for rotationally modulated light curves and surface magnetic fields in a few of the cooler subluminous B stars. 
\citet{momany20} reports extreme horizontal branch stars in globular clusters with $T_{\rm eff} = 20,000 - 30,000$ K having singly sinusoidal light curves. The preferred model requires a magnetic field of a few hundred Gauss and a single spot; pulsation is excluded. 
The only field sdB star in the paper, CD$-38^{\circ}222 = $SB\,290, has a similar light curve form in TESS data, which also show a flare. \citet{schneider18} report SB\,290 as one of a number of He$^3$ stars ($\Teff \approx 27000$ K), where strong vertical helium stratification is inferred.  

\subsection{Rossby waves}

 Rossby waves, or r-modes, are excited in the outer layers of rotating stars and planets. Their frequencies are predicted to lie in groups close to harmonics of the rotation frequency. 
Specifically, mode frequencies ($\nu_r$) lie in the range 
\[m \Omega[1 - 2/[(m+1)(m+2)] ] <  \nu_r < m \Omega \]
where the azimuthal wavenumber $m > 0$ 
and $\Omega \equiv f_{\rm rot}$ is the rotation frequency \citep{saio18b}. 

\citet{albekioni23} extended the theory of equatorially trapped Rossby waves, including results applicable to the radiative envelopes of hot stars. For HD 49798, adopting their first valid vertical mode gives a Lamb parameter \(\varepsilon \approx 2300\), placing the star firmly in the asymptotic regime where the mode frequencies approach integer multiples of the rotation frequency (their Fig. 4). 
 
\citet{jeffery20c} identified similar light variations in the helium-rich hot subdwarfs BD$+37^{\circ}442$ and BD$+37^{\circ}1977$. These stars show multiperiodic low-amplitude variability with principal periods of 0.56 and 1.14 d, respectively, with both first and second harmonics present. 
All three stars have $T_{\rm eff} \approx 45\,000 - 50\,000 $ K,  
 $\log g/{\rm cm\,s^{-2}} = 4.0 - 4.5$, and high helium / hydrogen surface abundances. 
Given these similarities, a Rossby wave interpretation provides a strong alternative explanation for the additional light variation in HD49798. 
The high harmonics visible in the power spectrum would then correspond with high azimuthal wavenumbers $m$, and the underlying 1.4\,d period would be close to or identical with the rotational period.

The TESS photometry does not distinguish between rotational modulation by long-lived surface structures and rotationally related Rossby-wave variability. Both predict variability tied to the stellar rotation period. Additional spectroscopy or spectropolarimetry covering the 1.4\,d cycle will be required to discriminate between them.

\section{Conclusion}
The TESS light curve of the X-ray binary HD\,49798 shows an ellipsoidal modulation at the orbital period  and an independent structured modulation folded on  or close to the rotation period inferred from spectroscopy. 
The first is certainly due to tidal deformation of the primary, whilst the latter  may arise from rotationally modulated surface structure, for example associated with magnetic activity or with Rossby (r-) modes in the stellar envelope. The observed frequencies for either explanation would be close to or identical with harmonics of the rotation frequency.

\section*{Acknowledgements}

The Armagh Observatory and Planetarium is core funded by the Northern Ireland Department for Communities.
This paper includes data collected with the TESS mission, obtained from the MAST data archive at the Space Telescope Science Institute (STScI). Funding for the TESS mission is provided by the NASA Explorer Program. STScI is operated by the Association of Universities for Research in Astronomy, Inc., under NASA contract NAS 5–26555. The authors thank the referee for constructive remarks, and Veronika Schaffenroth for sharing their draft paper on HD\,49798.  

\section*{Data Availability}

All data used in this paper are available from the MAST data archive.  



\bibliographystyle{mnras}
\bibliography{ehe} 

@article{albekioni23,
	adsurl = {https://ui.adsabs.harvard.edu/abs/2023A&A...671A..91A},
	archiveprefix = {arXiv},
	author = {{Albekioni}, M. and {Zaqarashvili}, T.~V. and {Kukhianidze}, V.},
	doi = {10.1051/0004-6361/202243985},
	eid = {A91},
	eprint = {2301.07446},
	journal = {\aap},
	month = mar,
	pages = {A91},
	primaryclass = {astro-ph.SR},
	title = {{Rossby waves on stellar equatorial {\ensuremath{\beta}} planes: Uniformly rotating radiative stars}},
	volume = {671},
	year = 2023}

@article{pelisoli22,
	adsurl = {https://ui.adsabs.harvard.edu/abs/2022MNRAS.515.2496P},
	archiveprefix = {arXiv},
	author = {{Pelisoli}, Ingrid and {Dorsch}, M. and {Heber}, U. and {G{\"a}nsicke}, B. and {Geier}, S. and {Kupfer}, T. and {N{\'e}meth}, P. and {Scaringi}, S. and {Schaffenroth}, V.},
	doi = {10.1093/mnras/stac1069},
	eprint = {2204.06575},
	journal = {\mnras},
	month = sep,
	number = {2},
	pages = {2496-2510},
	primaryclass = {astro-ph.SR},
	title = {{Discovery and analysis of three magnetic hot subdwarf stars: evidence for merger-induced magnetic fields}},
	volume = {515},
	year = 2022}

@article{mereghetti21,
	author = {Mereghetti, S and Pintore, F and Rauch, T and La Palombara, N and Esposito, P and Geier, S and Pelisoli, I and Rigoselli, M and Schaffenroth, V and Tiengo, A},
	copyright = {https://academic.oup.com/journals/pages/open\_access/funder\_policies/chorus/standard\_publication\_model},
	doi = {10.1093/mnras/stab1004},
	issn = {0035-8711, 1365-2966},
	journal = {Monthly Notices of the Royal Astronomical Society},
	language = {en},
	month = apr,
	number = {1},
	pages = {920--925},
	title = {New {X}-ray observations of the hot subdwarf binary {HD} 49798/{RX} {J0648}.0--4418},
	url = {https://academic.oup.com/mnras/article/504/1/920/6219849},
	urldate = {2026-05-27},
	volume = {504},
	year = {2021}}

@article{momany20,
	adsurl = {https://ui.adsabs.harvard.edu/abs/2020NatAs...4.1092M},
	archiveprefix = {arXiv},
	author = {{Momany}, Y. and {Zaggia}, S. and {Montalto}, M. and {Jones}, D. and {Boffin}, H.~M.~J. and {Cassisi}, S. and {Moni Bidin}, C. and {Gullieuszik}, M. and {Saviane}, I. and {Monaco}, L. and {Mason}, E. and {Girardi}, L. and {D'Orazi}, V. and {Piotto}, G. and {Milone}, A.~P. and {Lala}, H. and {Stetson}, P.~B. and {Beletsky}, Y.},
	doi = {10.1038/s41550-020-1113-4},
	eprint = {2006.02308},
	journal = {Nature Astronomy},
	month = jun,
	pages = {1092-1101},
	primaryclass = {astro-ph.SR},
	title = {{A plague of magnetic spots among the hot stars of globular clusters}},
	volume = {4},
	year = 2020}

@article{Ramiaramanantsoa14,
	adsurl = {https://ui.adsabs.harvard.edu/abs/2014MNRAS.441..910R},
	archiveprefix = {arXiv},
	author = {{Ramiaramanantsoa}, Tahina and {Moffat}, Anthony F.~J. and {Chen{\'e}}, Andr{\'e}-Nicolas and {Richardson}, Noel D. and {Henrichs}, Huib F. and {Desforges}, S{\'e}bastien and {Antoci}, Victoria and {Rowe}, Jason F. and {Matthews}, Jaymie M. and {Kuschnig}, Rainer and {Weiss}, Werner W. and {Sasselov}, Dimitar and {Rucinski}, Slavek M. and {Guenther}, David B.},
	doi = {10.1093/mnras/stu619},
	eprint = {1403.7843},
	journal = {\mnras},
	month = jun,
	number = {1},
	pages = {910-917},
	primaryclass = {astro-ph.SR},
	title = {{MOST detects corotating bright spots on the mid-O-type giant {\ensuremath{\xi}} Persei}},
	volume = {441},
	year = 2014}

@article{jackson12,
	adsurl = {https://ui.adsabs.harvard.edu/abs/2012ApJ...751..112J},
	archiveprefix = {arXiv},
	author = {{Jackson}, Brian K. and {Lewis}, Nikole K. and {Barnes}, Jason W. and {Drake Deming}, L. and {Showman}, Adam P. and {Fortney}, Jonathan J.},
	doi = {10.1088/0004-637X/751/2/112},
	eid = {112},
	eprint = {1203.6070},
	journal = {\apj},
	month = jun,
	number = {2},
	pages = {112},
	primaryclass = {astro-ph.EP},
	title = {{The EVIL-MC Model for Ellipsoidal Variations of Planet-hosting Stars and Applications to the HAT-P-7 System}},
	volume = {751},
	year = 2012}

@article{green23,
	author = {Green, Matthew J and Maoz, Dan and Mazeh, Tsevi and Faigler, Simchon and Shahaf, Sahar and Gomel, Roy and El-Badry, Kareem and Rix, Hans-Walter},
	copyright = {https://creativecommons.org/licenses/by/4.0/},
	doi = {10.1093/mnras/stad915},
	issn = {0035-8711, 1365-2966},
	journal = {Monthly Notices of the Royal Astronomical Society},
	language = {en},
	month = apr,
	number = {1},
	pages = {29--55},
	shorttitle = {15 000 ellipsoidal binary candidates in \textit{{TESS}}},
	title = {15 000 ellipsoidal binary candidates in \textit{{TESS}} : {Orbital} periods, binary fraction, and tertiary companions},
	url = {https://academic.oup.com/mnras/article/522/1/29/7100976},
	urldate = {2026-05-26},
	volume = {522},
	year = {2023}}

@article{schneider18,
	adsurl = {https://ui.adsabs.harvard.edu/abs/2018A&A...618A..86S},
	archiveprefix = {arXiv},
	author = {{Schneider}, D. and {Irrgang}, A. and {Heber}, U. and {Nieva}, M.~F. and {Przybilla}, N.},
	doi = {10.1051/0004-6361/201833182},
	eid = {A86},
	eprint = {1804.03234},
	journal = {\aap},
	month = oct,
	pages = {A86},
	primaryclass = {astro-ph.SR},
	title = {{NLTE spectroscopic analysis of the $^{3}$He anomaly in subluminous B-type stars}},
	volume = {618},
	year = 2018}

@article{rigoselli23,
	adsurl = {https://ui.adsabs.harvard.edu/abs/2023MNRAS.523.3043R},
	archiveprefix = {arXiv},
	author = {{Rigoselli}, Michela and {De Grandis}, Davide and {Mereghetti}, Sandro and {Malacaria}, Christian},
	doi = {10.1093/mnras/stad1611},
	eprint = {2305.15845},
	journal = {\mnras},
	month = aug,
	number = {2},
	pages = {3043-3048},
	primaryclass = {astro-ph.HE},
	title = {{Timing the X-ray pulsating companion of the hot subdwarf HD 49798 with NICER}},
	volume = {523},
	year = 2023}

@software{Pywhiten2023,
	author = {Stacey, Erik},
	title = {pywhiten: Lomb--Scargle-based prewhitening of stellar time series},
	url = {https://github.com/erikstacey/pywhiten},
	year = {2023}}

@article{Cantiello2011,
	adsurl = {https://ui.adsabs.harvard.edu/abs/2011A&A...534A.140C},
	archiveprefix = {arXiv},
	author = {{Cantiello}, M. and {Braithwaite}, J.},
	doi = {10.1051/0004-6361/201117512},
	eid = {A140},
	eprint = {1108.2030},
	journal = {\aap},
	month = oct,
	pages = {A140},
	primaryclass = {astro-ph.SR},
	title = {{Magnetic spots on hot massive stars}},
	volume = {534},
	year = 2011}

@article{rauch25,
	adsurl = {https://ui.adsabs.harvard.edu/abs/2025A&A...700A..80R},
	archiveprefix = {arXiv},
	author = {{Rauch}, T. and {Strau{\ss}}, P.},
	doi = {10.1051/0004-6361/202555181},
	eid = {A80},
	eprint = {2507.12912},
	journal = {\aap},
	month = aug,
	pages = {A80},
	primaryclass = {astro-ph.SR},
	title = {{Spectral analysis of HD 49798, a bright, hydrogen-deficient sdO-type donor star in an accreting X-ray binary}},
	volume = {700},
	year = 2025}

@article{Israel1996,
	author = {Israel, G. L. and Stella, L. and Angelini, L. and et al.},
	doi = {10.1086/310430},
	journal = {\apjl},
	pages = {L53--L56},
	title = {Discovery of 13.2 s X-Ray Pulsations from the Binary System HD 49798/RX J0648.0-4418},
	volume = {474},
	year = {1997}}

@article{mereghetti16b,
	author = {Mereghetti, S. and La Palombara, N. and Tiengo, A. and et al.},
	doi = {10.1093/mnras/stw556},
	journal = {\mnras},
	number = {4},
	pages = {3523--3531},
	title = {The puzzling compact object in HD 49798/RX J0648.0-4418},
	volume = {458},
	year = {2016}}

@article{Brooks2017,
	author = {Brooks, J. and Kupfer, T. and Bildsten, L. and et al.},
	doi = {10.3847/1538-4357/aa957d},
	journal = {\apj},
	pages = {105},
	title = {The Future Evolution of HD 49798},
	volume = {850},
	year = {2017}}

@article{Thackeray1970,
	author = {Thackeray, A. D.},
	journal = {\mnras},
	pages = {215--223},
	title = {The binary nature of HD 49798},
	volume = {150},
	year = {1970}}

@article{Landstreet2012,
	author = {Landstreet, J. D. and Bagnulo, S. and Fossati, L. and Jordan, S.},
	doi = {10.1051/0004-6361/201218681},
	journal = {\aap},
	pages = {A100},
	title = {A search for magnetic fields in hot subdwarf stars},
	volume = {541},
	year = {2012}}

@article{Ramiaramanantsoa2018,
	author = {Ramiaramanantsoa, T. and Moffat, A. F. J. and Harmon, R. and et al.},
	doi = {10.1093/mnras/stx2671},
	journal = {\mnras},
	number = {4},
	pages = {5532--5551},
	title = {BRITE-Constellation high-precision time-dependent photometry of the early O-type supergiant zeta Puppis unveils the photospheric drivers of its small- and large-scale wind structures},
	volume = {473},
	year = {2018}}

@article{jeffery20c,
	adsurl = {https://ui.adsabs.harvard.edu/abs/2020MNRAS.496..718J},
	archiveprefix = {arXiv},
	author = {{Jeffery}, C. Simon},
	doi = {10.1093/mnras/staa1555},
	eprint = {2006.00950},
	journal = {\mnras},
	month = jul,
	number = {1},
	pages = {718-722},
	primaryclass = {astro-ph.SR},
	title = {{TESS photometry of helium-rich hot subdwarfs: r modes in BD+37{\textdegree}442 and BD+37{\textdegree}1977}},
	volume = {496},
	year = 2020}

@article{dorsch24,
	adsurl = {https://ui.adsabs.harvard.edu/abs/2024A&A...691A.165D},
	archiveprefix = {arXiv},
	author = {{Dorsch}, M. and {Jeffery}, C.~S. and {Philip Monai}, A. and {Tout}, C.~A. and {Snowdon}, E.~J. and {Monageng}, I. and {Scott}, L.~J.~A. and {Miszalski}, B. and {Woolf}, V.~M.},
	doi = {10.1051/0004-6361/202451306},
	eid = {A165},
	eprint = {2410.02737},
	journal = {\aap},
	month = nov,
	pages = {A165},
	primaryclass = {astro-ph.SR},
	title = {{Discovery of three magnetic helium-rich hot subdwarfs with SALT}},
	volume = {691},
	year = 2024}

@article{snowdon25,
	adsurl = {https://ui.adsabs.harvard.edu/abs/2025MNRAS.537.2079S},
	archiveprefix = {arXiv},
	author = {{Snowdon}, E.~J. and {Jeffery}, C.~S. and {Schlagenhauf}, S. and {Dorsch}, M.},
	doi = {10.1093/mnras/staf152},
	eprint = {2501.13656},
	journal = {\mnras},
	month = feb,
	number = {2},
	pages = {2079-2089},
	primaryclass = {astro-ph.SR},
	title = {{A search for close binary systems in the SALT survey of hydrogen-deficient stars using TESS}},
	volume = {537},
	year = 2025}

@article{barlow22,
	adsurl = {https://ui.adsabs.harvard.edu/abs/2022ApJ...928...20B},
	archiveprefix = {arXiv},
	author = {{Barlow}, Brad N. and {Corcoran}, Kyle A. and {Parker}, Isabelle M. and {Kupfer}, Thomas and {N{\'e}meth}, P{\'e}ter and {Hermes}, J.~J. and {Lopez}, Isaac D. and {Frondorf}, Will J. and {Vestal}, David and {Holden}, Jazzmyn},
	doi = {10.3847/1538-4357/ac49f1},
	eid = {20},
	eprint = {2112.11463},
	journal = {\apj},
	month = mar,
	number = {1},
	pages = {20},
	primaryclass = {astro-ph.SR},
	title = {{New Variable Hot Subdwarf Stars Identified from Anomalous Gaia Flux Errors, Observed by TESS, and Classified via Fourier Diagnostics}},
	volume = {928},
	year = 2022}

@article{snowdon23b,
	adsurl = {https://ui.adsabs.harvard.edu/abs/2023MNRAS.525..183S},
	author = {{Snowdon}, E.~J. and {Jeffery}, C.~S. and {Schlagenhauf}, S. and {Dorsch}, M. and {Monageng}, I.~M.},
	doi = {10.1093/mnras/stad2303},
	journal = {\mnras},
	month = oct,
	number = {1},
	pages = {183-189},
	title = {{Ton S 415: a close binary containing an intermediate helium subdwarf discovered with SALT and TESS}},
	volume = {525},
	year = 2023}

@misc{lightkurve18,
	adsurl = {http://adsabs.harvard.edu/abs/2018ascl.soft12013L},
	archiveprefix = {ascl},
	author = {{Lightkurve Collaboration} and {Cardoso}, J.~V.~d.~M. and {Hedges}, C. and {Gully-Santiago}, M. and {Saunders}, N. and {Cody}, A.~M. and {Barclay}, T. and {Hall}, O. and {Sagear}, S. and {Turtelboom}, E. and {Zhang}, J. and {Tzanidakis}, A. and {Mighell}, K. and {Coughlin}, J. and {Bell}, K. and {Berta-Thompson}, Z. and {Williams}, P. and {Dotson}, J. and {Barentsen}, G.},
	eprint = {1812.013},
	howpublished = {Astrophysics Source Code Library},
	month = dec,
	title = {{Lightkurve: Kepler and TESS time series analysis in Python}},
	year = 2018}

@article{dorsch22,
	adsurl = {https://ui.adsabs.harvard.edu/abs/2022A&A...658L...9D},
	archiveprefix = {arXiv},
	author = {{Dorsch}, M. and {Reindl}, N. and {Pelisoli}, I. and {Heber}, U. and {Geier}, S. and {Istrate}, A.~G. and {Justham}, S.},
	doi = {10.1051/0004-6361/202142880},
	eid = {L9},
	eprint = {2201.08146},
	journal = {\aap},
	month = feb,
	pages = {L9},
	primaryclass = {astro-ph.SR},
	title = {{Discovery of a highly magnetic He-sdO star from a double-degenerate binary merger}},
	volume = {658},
	year = 2022}

@article{saio18b,
	adsurl = {https://ui.adsabs.harvard.edu/abs/2018arXiv181201253S},
	archiveprefix = {arXiv},
	author = {{Saio}, Hideyuki},
	eid = {arXiv:1812.01253},
	eprint = {1812.01253},
	journal = {arXiv e-prints},
	month = dec,
	pages = {arXiv:1812.01253},
	primaryclass = {astro-ph.SR},
	title = {{R mode oscillations ubiquitous in stars}},
	year = 2018}

@article{ahmad05b,
	adsurl = {http://ukads.nottingham.ac.uk/abs/2005A%26A...437L..51A},
	author = {{Ahmad}, A. and {Jeffery}, C.~S.},
	journal = {\aap},
	month = jul,
	pages = {L51-L54},
	title = {{Discovery of pulsation in a helium-rich subdwarf B star}},
	volume = 437,
	year = 2005}

@article{randall11,
	adsurl = {http://adsabs.harvard.edu/abs/2011ApJ...737L..27R},
	author = {{Randall}, S.~K. and {Calamida}, A. and {Fontaine}, G. and {Bono}, G. and {Brassard}, P.},
	doi = {10.1088/2041-8205/737/2/L27},
	eid = {L27},
	journal = {\apjl},
	month = aug,
	pages = {L27},
	title = {{Rapidly Pulsating Hot Subdwarfs in {$\omega$} Centauri: A New Instability Strip on the Extreme Horizontal Branch?}},
	volume = 737,
	year = 2011}

@article{landstreet78,
	adsurl = {http://cdsads.u-strasbg.fr/abs/1978ApJ...224L...5L},
	author = {{Landstreet}, J.~D. and {Borra}, E.~F.},
	doi = {10.1086/182746},
	journal = {\apjl},
	month = aug,
	pages = {L5-L8},
	title = {{The magnetic field of Sigma Orionis E}},
	volume = 224,
	year = 1978}

@article{koen97,
	adsurl = {http://adsabs.harvard.edu/abs/1997MNRAS.285..645K},
	author = {{Koen}, C. and {Kilkenny}, D. and {O'Donoghue}, D. and {van Wyk}, F. and {Stobie}, R.~S.},
	journal = {\mnras},
	month = mar,
	pages = {645-650},
	title = {{A new class of rapidly pulsating star - II. PB 8783}},
	volume = 285,
	year = 1997}

@article{woudt06,
	adsurl = {http://adsabs.harvard.edu/abs/2006MNRAS.371.1497W},
	author = {{Woudt}, P.~A. and {Kilkenny}, D. and {Zietsman}, E. and {Warner}, B. and {Loaring}, N.~S. and {Copley}, C. and {Kniazev}, A. and {V{\"a}is{\"a}nen}, P. and {Still}, M. and {Stobie}, R.~S. and {Burgh}, E.~B. and {Nordsieck}, K.~H. and {Percival}, J.~W. and {O'Donoghue}, D. and {Buckley}, D.~A.~H.},
	doi = {10.1111/j.1365-2966.2006.10788.x},
	eprint = {arXiv:astro-ph/0607171},
	journal = {\mnras},
	month = sep,
	pages = {1497-1502},
	title = {{SDSS J160043.6+074802.9: a very rapid sdO pulsator}},
	volume = 371,
	year = 2006}

@article{kilkenny97,
	adsurl = {http://adsabs.harvard.edu/abs/1997MNRAS.287..867K},
	author = {{Kilkenny}, D. and {O'Donoghue}, D. and {Koen}, C. and {Stobie}, R.~S. and {Chen}, A.},
	journal = {\mnras},
	month = jun,
	pages = {867-893},
	title = {{The Edinburgh-Cape Blue Object Survey - II. Zone 1 - the North Galactic CAP}},
	volume = 287,
	year = 1997}

\bsp	
\label{lastpage}
\end{document}